\documentclass{JHEP3}
\usepackage{epsf}
\usepackage{amsmath}
\usepackage{amsfonts}
\usepackage{amssymb}

\newcommand{\pa}{\partial}
\newcommand{\tr}{{\rm tr}}
\newcommand{\comment}[1]{}

\newcommand{\pasl}{\pa\kern-.55em /}

\newcommand{\ksl}{k\kern-.55em /}

\DeclareFixedFont{\xiiss}{OT1}{cmss}{m}{n}{12}
\DeclareFixedFont{\ixss}{OT1}{cmss}{m}{n}{9}
\DeclareFixedFont{\cmrnine}{OT1}{cmr}{m}{n}{9}
\newcommand{\field}[1]{\mathbb{#1}}
\newcommand{\BC}{{\field C}}

\newcommand{\BZ}{{\field Z}}

\newcommand{\CCs}{\hbox{\ixss C\kern-.4emI}}
\newcommand{\ZZs}{\hbox{\ixss Z\kern-.4emZ}}

\newcommand{\CM}{{\cal M}}

\newcommand{\CP}{{\BC\field P}}

 \newcommand{\myfig}[3]{\begin{figure}[ht]
\begin{center}
\leavevmode
\epsfxsize=#2cm
\epsfbox{#1}
\end{center}
\caption{#3}
\label{fig:#1}
\end{figure}}

\title{On the universality class of the conifold}

\author{David Berenstein \\
School of Natural Sciences,
Institue of Advanced Study,
Einstein Drive, NJ 08540\\
Email: \email{dberens@ias.edu} }

\preprint{hep-th/0110184\\NSF-ITP-01-161
}
\abstract{The possibility of having 
discrete degrees of freedom  at singularities
associated to `conifolds with discrete torsion' is studied.
We find that the field theory of D-brane probes near
these singularities is identical to ordinary 
conifolds, so that there are no additional discrete degrees
of freedom located at the singularity. We  shed light on 
how the obstructions to resolving the singularity are global
topological issues rather that local obstrucions at the singularity itself.
We also analyze the geometric transitions and duality cascades
when one has fractional branes 
at the singularity and compute the moduli space of an arbitrary
 number of probes in the geometry. 
We provide some evidence for a conjecture
that there are  no discrete degrees of freedom localized 
at any Calabi-Yau singularity that can not be guessed from topological data
 away from the singularity.
}

\begin{document}

It was proposed in \cite{VW} that there are conifold singularities in 
Calabi-Yau
manifolds that can not be resolved or deformed. These singularities
arise from deforming the
complex structure of a $T^6/\BZ_2\times \BZ_2$ orbifold with discrete torsion
by marginal deformations in the string wsorldsheet.
As such, it seems that these singularities are new classes of conifolds.
For this problem, it is important to 
understand if this obstruction is a property of the singularity itself, or
instead a topological obstruction from the embedding of the singularity
in a given Calabi-Yau space. Since the worldsheet conformal field theory is 
non-singular (one can obtain the conifold singularity with an infinitesimal
deformation away from the orbifold point) one can use D-branes as probes of
the geometry without the 
fear of them becoming massless at the singularity.

If we consider a large volume Calabi-Yau space 
(we make the orbifold non-compact) we can study the orbifold geometry
 point with
a collection of D3-branes. This gives us a probe of the geometry associated 
to the orbifold, and produces a near horizon geometry which is dual to a 
four dimensional CFT \cite{M,KW, MP}.
The rules for calculating correlations functions in the field theory and how
 to compare them to gravity were laid out in \cite{W,GKP}. 
On this orbifold geometry, one can find 
superpotential deformations of the field theory which correspond to 
infinitesimal deformations of the complex structure of the Calabi-Yau 
space \cite{D,DF}, and which lead to a conifold singularity 
in the moduli space of 
D-branes.

If this `conifold with discrete torsion'
 is topologically distinct from the standard conifold, then one
can compare two distinct field theories that realize the singularity 
as an infrared fixed point of the RG flow to decide this question.
 These two field theories are given by
the deformed field theory with discrete torsion near the point in
the moduli space of vacua
where the singularity is located; and the field theory associated to the
 conifold as realized by Klebanov and Witten \cite{KW}.
The test will be to see if these two a priori distinct
 field theories belong to the
 same universality class of field theories or not. 

Orbifold type singularities can be readily understood by the quiver 
diagram techniques of
Douglas and Moore \cite{DM}. Experience with these examples has proven
 that one can in general expect that all of the information of the singularity 
can be encoded in a four dimensional conformal field theory, that bulk
D-branes fractionate at the singularity and
 that
the moduli space of D-branes will recover the algebraic geometry of the 
singularity.  

The next simplest singularity is the conifold, which has been the subject of
 a lot of attention through the years, as it presents us with topology 
changing transitions \cite{S,GMS}, interacting conformal  field theories 
which do not have a free field theory limit \cite{KW}, and because it
 allows us to study confining field theories. These give
 rise to topology changing 
transitions \cite{KS,GV,CIV}. Brane realizations of these theories have been 
considered in \cite{DOT,DOT2,DOT3}.

One needs a systematic approach to construct these field theories and to 
interpret how brane fractionation occurs.
In \cite{BJL, BL4} it was proposed to study the (perturbative)
moduli space 
of these theories in terms of the representation theory of  non-commutative 
algebras. The possibility of 
fractional branes can then be understood in terms of families of
irreducible representations of the algebras which, for certain values of the 
parameters, 
become reducible. At these points the D-branes split into smaller
 irreducible 
representations. This should be the signal for the presence of singularities.
From the knowledge of the algebra and the irreducible
representations, one can compute explicitly the 
quiver diagram of the singularity \cite{BL4,BG} and obtain results 
which give exactly the quiver diagrams of Douglas and Moore \cite{DM},
but which are also applicable to non-orbifold singularities, once the 
noncommutative algebra is known. Moreover, it has
 been shown that one can build compact Calabi-Yau spaces which correspond
to  non-commutative versions of algebraic geometry \cite{BL3}.
Naturally, since we are analyzing singularities in a commutative 
geometry, one needs a way to understand this commutative structure from 
the noncommutative
point of view.  This commutative structure is provided by the center of the
 algebra.

Once one understands that fractional branes can arise, one can study
how the theories get affected by the presence of 
fractional branes (if they are allowed by anomaly cancellation),
and we should generally expect  that  these
fractional brane constructions lead to field theories that are
closely associated to the four dimensional conformal field 
theory of the singularity-- but which are no longer 
conformal. These associated field theories are expected
to have all kinds of interesting 
nonperturbative phenomena like confinement and duality cascades
\cite{KS}.

In principle, given a singularity it is unclear whether this collection
of  associated field theories is unique. 
There could be more than one algebra
\footnote{Properly we should talk of  equivalence classes of algebras}
 which gives rise to the same type of singular geometry.
The aim of this paper is to study this problem, namely the
question of whether the field theories at singularities in Calabi-Yau 
threefolds are universal or if there are extra
 discrete degrees of 
freedom in string theory which are necessary to specify such a class
of theories.
Most of our work will be done for the conifold singularity
and we will find that, at least in the example of `conifolds with 
discrete torsion', they are described by the same universality class of the 
conifold without discrete torsion, but with some relations between the gauge
couplings fixed from boundary conditions in the UV.

Towards the end of the paper we give a
holographic argument which supports the idea that the quantum
 field theories  of D-branes
on any  Calabi-Yau singularity are indeed universal
(up to gauge field theory dualities \cite{Sei}). The singularities
are allowed to 
have discrete torsion degrees of freedom only if the singularity 
is not isolated. This extra data is encoded 
by the monodromies of 
fractional branes of the codimension two singularities around the
codimension three singularities. As such, these monodromies 
correspond to 
holographic data away from the singularity and are specifying the
'stringy' topological type of the singularity. However, our point of
view will be that this data is not concentrated at the singularity itself,
since it can be measured far away from the singularity.
Given this 
additional data,
this should determine completely the universality class of field theories 
at the singularity (both with and without fractional branes) and the duality
cascades between different realizations of the field theory. These field 
theories associated to the singularity are only given by the infrared 
fixed point of a collection of D-branes very near the singularity, and in the
 presence
of fractional branes these may produce 
topology changing transitions and 
confinement.

The paper is organized as follows:

Section \ref{sec:algebra} describes the conifold
 algebra in detail using the techniques of \cite{BJL,BL4} and
the moduli space of vacua is shown to be the symmetric product of the 
conifold geometry.

Section \ref{sec:dt}  presents a field theory which corresponds to 
the orbifold with discrete torsion, and a deformation which leads to a 
conifold singularity. We expand the field theory about a background of 
Dbranes near the singularity and we show that the IR fixed point of the 
theory gives rise to the same field theory as the one studied in 
\cite{KW}, which corresponds to the conifold without discrete torsion.

Section \ref{sec:md} deals with the reasons why certain marginal
 deformations of the conifold geometry can be obstructed in the field
theory with discrete torsion. We find a consistent picture from various 
points of view which solve the puzzle presented by \cite{VW}.

In section \ref{sec:cascade} we further advance the geometric picture of geometric transitions 
and the understanding of Seiberg dualities. In particular,
we are able to compute the  moduli space of an arbitrary number of
D3-brane probes in the presence of fractional branes including 
the non-perturbative superpotential, and we find that it is 
indeed given by the symmetric product of the deformed conifold.

In section \ref{sec:hol} we speculate on universality for more general 
singularities and  propose a holographic
argument that gives evidence for any Calabi-Yau singularity to behave 
universally.

The paper is concluded in section \ref{sec:con}.

\section{D-branes and the conifold algebra}\label{sec:algebra}

The conifold geometry can be described by the variety which 
is the solution to the equation
\begin{equation}
a^2+b^2+c^2+d^2 =0
\end{equation}
in four variables, and represents a conical singularity which is
also a Calabi-Yau manifold.

As the geometry of the singularity is a cone, one also expects
that setting a collection of D-branes at the singularity will produce
a conformal field theory \cite{KW,MP}. The 
cone is over the homogeneous space $(SU(2)\times SU(2))/U(1)$. The $U(1)$ is 
embedded diagonally. The group manifold has an $SU(2)^4$ isometry group 
which is broken to $SU(2)\times SU(2)\times U(1)$
upon taking the quotient. In the supergravity theory
 the $U(1)$ direction will be interpreted as the R-charge of the conformal 
symmetry group.

A D-brane probe on the geometry should have the conifold as a moduli 
space of vacua.
In general, one can think of having $n$ D-brane probes of 
the geometry,
and it is expected that
the classical moduli space of such probes should give us 
the $n$th symmetric product of the conifold
geometry. 

A change of variables lets us write the conifold as
\begin{equation}
uv = w z
\end{equation}
so that the conifold geometry can  also be constructed
 as the holomorphic quotient
$\BC^4/\BC^*$ with charges $[-1,-1,1,1]$ on variables $a_1, a_2, b_1, b_2$.
The invariants under such an action are the four  $a_i b_j$, which are 
identified with $u,v,w,z$.

We can also rewrite the conifold geometry as
\begin{equation}\label{eq:def1}
uv = (w'-\lambda \gamma)(w'+\lambda\gamma)
\end{equation}
with a parameter $\lambda$. We can set later $\lambda=1$
so that we have a family of conifold singularities, which at $\lambda =0$
gives us the orbifold space $(\BC^2/\BZ_2)\times \BC$. Thus, one can obtain
 the conifold geometry as a deformation of the complex structure of the 
orbifold $(\BC^2/\BZ_2)\times \BC$.

For  D-branes, this means  that one should be able to obtain the
field theory of D-branes on a conifold geometry from a superpotential
deformation of the supersymmetric 
field theory associated to the above orbifold, and then take the infrared
limit near the conifold singularity in moduli space.

The approach of Klebanov and Witten \cite{KW}
 to the conifold field theory  is given 
by a holomorphic quotient $(\BC^4)/\BC^*$.
The theory has four superfields
$x_1,x_2, y_1, y_2$ with charges $(1,1,-1,-1)$ under a $U(1)$ gauge field, 
and with no superpotential. There is obviously an $SU(2)\times SU(2)$
symmetry which acts on the $x_i$ and $y_i$ as $(2,1)$ and $(1,2)$ 
respectively.

The moduli space is given by the vevs of the gauge invariant
superfields 
\begin{equation}\label{eq:Nij}
N_{ij}= x_i y_j
\end{equation}
 and one finds the classical constraint in terms of the $2\times 2$
matrices
$\det(N_{ij})=0$, which gives rise to the conifold geometry.

To make further contact with D-branes, one has to 
interpret how the field theory is affected when one adds $N$ 
branes. Clearly  one has to extend the gauge group to 
$U(N)$ somehow and one wants to be able to break $U(N)\to U(1)^N$
by classical vevs of $x_i, y_i$. If the $x_i$ are in the $N$ and
the $y_i$ are in the $\bar N$ of $U(N)$,
this means that we need $N$
of the $x_i$ and $y_i$ in order to get $N$ distinct points on the 
conifold.  This would lead to an additional 
$U(N)$ symmetry of the theory which is not visible in the supergravity,
so this symmetry must be gauged. The $U(N)\times U(N)$ is expected from 
D-brane gauge groups if there are two types of fractional branes.
 One linear combination of the diagonal
$U(1)'s$ is decoupled (no field is charged under it),
while under the other $U(1)$ the fields $x,y$ have opposite charges.

The end result is a field theory which can be represented by the
 quiver diagram drawn in figure \ref{fig:quiver.eps}

\myfig {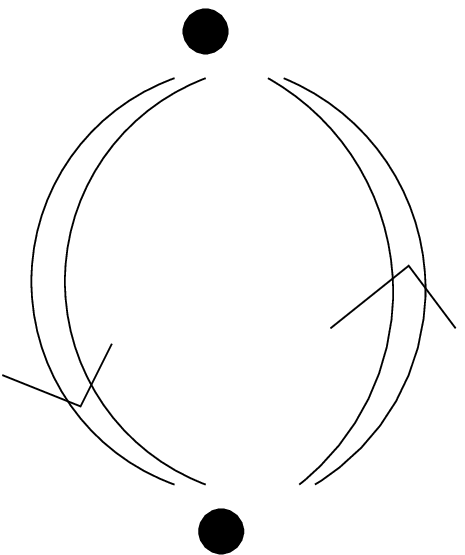}{3}{Quiver diagram for the conifold field theory}

One also needs to add a superpotential so that the
 moduli space of vacua of the theory is of dimension 
$N$ rather than of order $N^2$ (the number of free fields that make 
the theory). If one requires the additional symmetry under the exchange of 
the two gauge groups then one can argue that the fields $x_i$ and $y_i$ 
 have the same dimension which has to be 
set equal to $\dim(x_i)=3/4$. The $R$ charges of the fields
 are fixed by the requirement of
 conformal invariance.
The necessary superpotential deformation
is quartic and unique given the $SU(2)\times SU(2)$ global symmetry
and being of single trace type 
\begin{equation}
W= \lambda \tr(N_{ij} N_{lk} \epsilon_{il}\epsilon_{jk})
\end{equation}
with $N_{ik} = x_i y_k$.
This theory can be shown to be conformal by using the 
Leigh-Strassler techniques \cite{LS}, as the $\beta$ functions
for the gauge couplings and the superpotential are related.

To make contact with the algebraic techniques of \cite{BJL},
we can also write this superpotential in terms of fields $X$, $Y$
which are given by the $2N\times 2N$ matrices
\begin{equation}\label{eq:matrices}
X = \begin{pmatrix}
0 & x_1\\
y_1 & 0
\end{pmatrix},\quad
Y =  \begin{pmatrix}
0 & x_2\\
y_2 & 0
\end{pmatrix}, \sigma= \begin{pmatrix}
1&0\\
0 & - 1
\end{pmatrix}
\end{equation}
so that 
\begin{equation}\label{eq:sup1}
W = \lambda\tr(\sigma[X,Y]^2)
\end{equation}
The reason for introducing the discrete variable $\sigma$ is as
a placeholder for the $U(1)$ gauge group in the $U(N)\times U(N)$
theory. In the case of one brane,
it is exactly gauging by $\exp(\theta\sigma)$ which gives rise to the
correct $U(1)$ action on $X,Y$, and it is easy to check that
the above superpotential gives zero. In terms of the 
$2N\times 2N$ matrices the $U(N)\times U(N)$
gauge group is selected by requiring that it commute with 
$\sigma$.

We also note that 
\begin{equation}
X \sigma  = -\sigma X, Y \sigma  = -\sigma Y, \sigma^2 =1\label{eq:rel1}
\end{equation} 

There are also
constraints derived from the superpotential which read
\begin{equation}
(\sigma[X,Y], X) = (\sigma[X,Y], Y) = 0\label{eq:rel2}
\end{equation}

We will call the  polynomial algebra generated by $X,Y ,\sigma$--
subject to the relations 
\ref{eq:rel1} and \ref{eq:rel2} -- the algebra of the conifold.

Let us make one last note as to how we can obtain this field theory from 
the deformation of complex structure of the $\BC^2/\BZ_2\times \BC$
algebra. From equation \ref{eq:def1} we see that we get a family
of resolved ADE singularities, with a deformation that depends
on the third coordinate $\gamma$.  These deformations of complex structure are
given by twisted fields in the worldsheet conformal field theory, and
this translates into $F$ terms for the adjoints in the 
quiver of the $\BC^2/\BZ_2$
${\cal N} = 2$ quantum field theory \cite{DGM}, all we need now is to make them
 dependent on
 the third coordinate.
The ${\cal N}=2$ $A_1$ theory is described by the quiver diagram
in figure
\ref{fig:quiver3.eps}.

\myfig{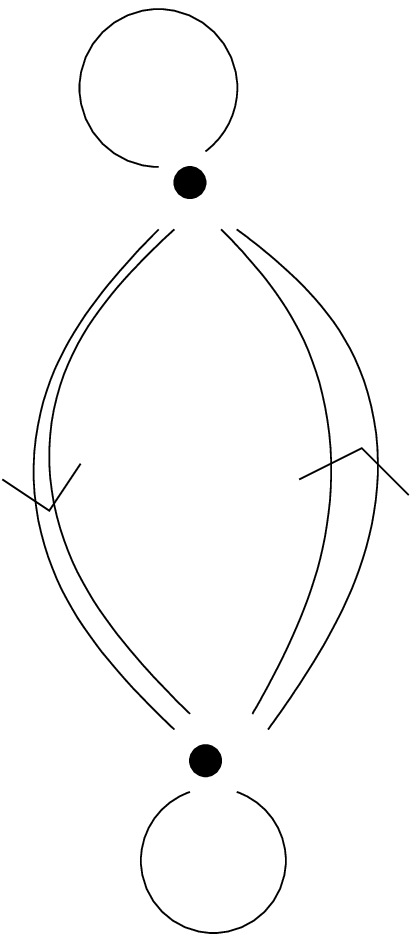}{3}{Quiver diagram for the $A_1$ singularity}

The associated noncommutative
algebra is given by the crossed product algebra of $\BC^3$ and the group
algebra $\BZ_2$ \cite{BL4}. Let $\sigma$ be the $\BZ_2$
group variable, thus $\sigma^2=1$
\begin{eqnarray}
\sigma X  &=& -  X\sigma \\
\sigma Y &=& - Y\sigma\\
\sigma Z &=& - Z\sigma
\end{eqnarray}
The superpotential is given by $\tr([X,Y],Z)$, the same as the 
${\cal N}= 4$ field theory, which leads to $X,Y, Z$ commuting with each other.
 All we have done is added the extra
discrete variable 
$\sigma$, which projects us onto the $\BZ_2$ invariant states. 

Now we need to perturb the theory by a twisted chiral field. These are
of the form \footnote{This is a rewriting of the results of 
\cite{BL} in these matrix variables.}
\begin{equation}
-\frac 1n \tr(\sigma Z^n).
\end{equation}
They only modify the fact that  $X,Y$ commute, so we get
to
\begin{equation}
[X,Y] = \sigma Z^{n-1}
\end{equation}
From here, the center of the algebra 
is always given by $Z,u=  X^2,v= Y^2, \gamma= XY+YX$, and the 
relation between these commutative 
variables changes because when we evaluate
$\gamma^2$ we need to commute $X,Y$ to order them. In this way,
\begin{equation}
\gamma^2 = 4 u^2+ Z^{2(n-1)}. 
\end{equation}
We need $n=2$ to obtain the conifold. In this case
$Z$ has a mass term, so it can be integrated out,
 and the only surviving
generators are $X,Y$ with an induced superpotential given by
\begin{equation}
\tr(\sigma([X,Y])^2)
\end{equation}
which again reproduces equation \ref{eq:sup1}. As the field $Z$ is 
integrated out, it 
dissapears from the quiver diagram and we are left with figure 
\ref{fig:quiver.eps} instead.

\subsection{On the  fate of the  $U(1)$'s}

So far we have described the field theory obtained in the infrared as
 the fixed point
of the RG flow from a $U(N)\times U(N)$ field theory in the UV which 
is matched to a string background. 

In particular, one describes the UV boundary conditions of the flow by
 the gauge couplings of the $U(N)^2$ theory. There is one $U(1)$ which, as far
as the low energy field theory degrees of freedom are concerned, is decoupled
from the open string sector because no field is charged under it. This field
 participates in the gauge variation of 
the closed string 
$B$ field and can be gauged away.
If the closed string fields decouple from the low energy degrees
of freedom then this global $U(1)$ is not part of the low energy dynamics on
 the brane at all.

There is a second  $U(1)$ term in the Lagrangian which is the relative $U(1)$
(the difference of the two diagonal
$U(1)$ charges in the $U(N)\times U(N)$ theory). 
This term  allows us to give a D-term to the Lagrangian, and there
 is matter 
coupled to this $U(1)$. At the string scale (or some other higher energy
 scale where we are
 matching the field theory),
this gauge coupling is determined by the two 
$SU(N)$ couplings, but as we flow towards the infrared the running
 gauge coupling goes to zero
from screening by the massless charged fields.
At the infrared fixed point  theory the $U(1)$ is decoupled. When we
 describe the 
infrared field theory by an AdS background, this flow towards the infrared
 has already
been taken into account, so in the conformal field theory the $U(1)$ is not
 properly there
 any more. If we are on a resolved conifold, the D-term of this $U(1)$ 
is part of the low energy data, and one has to keep a scaling limit where the
 coupling of
the $U(1)$ goes to zero, but where the effects of the D-term give a finite 
scale.
However, these massless fields are not in the spectrum of the conformal field
 theory 
associated to the AdS background.

\subsection{Representations of the conifold algebra}

Now we want to understand the classical moduli spaces of vacua of the 
conifold theory. We follow the techniques presented in \cite{BJL,BL4}.

From the algebraic relations we find that 
\begin{equation}
u= X^2, v=  Y^2
,w= (XY+YX),z=  [X Y - YX]\sigma
\end{equation}
all belong to the center of the algebra, and thus they describe 
a commutative geometry. One can easily show that
\begin{equation}
4 u v = w^2 - z^2 
\end{equation}
which, up to some trivial rescalings, gives us the commutative
conifold geometry.
On each irreducible representation of the algebra we have that
all of $u,v,w,z$ are proportional to the identity, and thus specify a 
single point in the conifold. It is a simple exercise to show that, 
away from the singularity, there is one unique irreducible representation
of the algebra in terms of $2\times2$ matrices
for each of the conifold points, and they are constructed by
giving arbitrary values to the variables $x_i, y_i$
in equations \ref{eq:matrices}, subject to the gauge
equivalence generated by $\exp\theta\sigma$ (which changes the values
of the parameters). Each of these representations is such that
$\tr(\sigma)=0$, which means that it is built out of two fractional branes, 
one of each type.

At the conifold singularity, we find additional irreducible representations
of the algebra which are one-dimensional
and  which are given by taking $x_i= y_i =0$, as  all
the equations are satisfied automatically except $\sigma^2= 1$. 
The two possible solutions correspond to $\sigma = \pm 1$, which
we will call $R_-$ and $R_+$. If we have $N R_-$ and $M R_+$ branes, then the
 gauge group is $U(N)\times U(M)$.

It is easy to see that the bulk representation decomposes as
$R_{bulk} \to R_-\oplus R_+$. These are the fractional branes.
They are characterized by the distinct values of 
$\tr(\sigma) = \pm 1$, so we can use $\tr(\sigma)$ as the parameter
that counts how many fractional branes we have.

The end result is that branes only fractionate at the singularity. There
are two types of fractional branes and a bulk brane fractionates into
one of each of the two types of fractional branes. The gauge group of a 
configuration of $N$ bulk branes with $N-M$ fractional branes is the 
gauge group associated to $N$ and $M$ fractional
branes of different kind, namely $U(N)\times U(M)$.

At least in principle, the moduli space of the theory is given by a
 direct sum
 of irreducible
representations of the algebra, and as such it is identified with the
a set of $N$ points in the conifold.

\subsection{The resolved conifold}

Although we have written the holomorphic quotient na\"\i vely as 
$\BC^4/\BC^*$ one would really want to be more careful and
write the quotient as a symplectic quotient to take into account the
metric aspects of the singularity.

If we do this, one needs to consider also the D-term equations of
motion of the supersymmetric field theory.

In the conifold field theory
language, the D-term constraints are given by
\begin{eqnarray}\label{eq:Dterms}
[\bar X, X] + [\bar Y, Y] = D\sigma
\end{eqnarray}
or equivalently
\begin{equation}
|x_i|^2 - |y_i|^2 = D
\end{equation}
where we have introduced complex conjugate coordinates for $X, Y$ (which are
 their adjoints in the matrix algebra and give the algebra the structure of a 
$\BC^*$ algebra).
Since we have that $X\bar X$ is a gauge invariant function in
$\BC^4$, it descends to an element of the center in the 
quotient algebra, and
similarly for $Y\bar Y$. On the moduli space of vacua
the equations
\ref{eq:Dterms} are block diagonal for the individual branes.
Thus we only need to analyze one bulk brane at a time.
All of the individual D-terms will be identical to one another as the 
supersymmetric field theory can only accommodate non-zero D-terms for the 
non-decoupled $U(1)$ in the $SU(N)\times U(1) \times SU(N)\times U(1)$ field 
theory.

We already have representations of $X,Y$
 in terms of $2\times 2$ matrices as described in equations \ref{eq:matrices},
 and we take $\bar X, \bar Y$ to be the 
adjoints in this representation. The conifold is described  by 
the level set $D=0$. In this case, the 
representations at the origin are those for which 
$x_i y_j=0$ so if we assume 
one of the $x_i\neq 0$ then both of the $y_j$ are zero
and from \ref{eq:Dterms} we find that $|x|^2 = 0$ as well.
In this case the holomorphic  branes truly fractionate at the origin.

For distinct level sets, e.g. $D>0$, we have that 
$|x|^2 >0$, so the origin of the conifold in moduli space has
a nonzero value for (at least one of) the coordinates $x_1, x_2$.
The ratio 
\begin{equation}\label{eq:blowup}
\kappa_{ij} = x_i/x_j= N_{ij}/N_{jj}
\end{equation}
is a 
holomorphic invariant for  this 
configuration so long as the denominator is nonzero. This requires
us to split the coordinates in moduli space in two patches
where $x_{1,2}\neq 0$ and $y=0$.

As such, when we include these two patches we have a $\CP^1$
space, as $\kappa_{12}\kappa_{21} = 1$ when
 both are defined.
It is also clear that the interpretation of the origin in moduli space 
is that one has  blown-up the singularity to a finite
size $\CP^1$ as follows from the description in terms of quotients
of commutative variables in equation \ref{eq:blowup}.

Choosing $D<0$ exchanges the role of $x,y$ and corresponds to
flopping the rigid $\CP^1$.

From the holomorphic quotient point of view, these two blow-ups can be 
interpreted as holomorphic quotients
\begin{equation}
(\BC^4/E)/\BC^*
\end{equation}
where we remove one of two exceptional sets:
either the complex variety $y_i=0$ or the complex variety $x_i=0$
so that irreducible representations with these values of the
complex parameters are not allowed. 
Presented in this manner, there is no need to specify values for the 
D-terms and one is only using algebraic geometric 
descriptions of the blow-ups.

\section{The conifold with discrete torsion}{\label{sec:dt}}

One can also obtain a conifold singularity from a deformation of the complex
structure of the $T^6/\BZ_2\times\BZ_2$ orbifold with
 discrete torsion \cite{VW}.

 A noncompact version of this space is 
a $\BC^3/\BZ_2\times \BZ_2$ with discrete torsion.
The $\BZ_2\times \BZ_2$ orbifold is given by the variety described by
 the
 equations
\begin{equation}
\gamma^2 = u v w
\end{equation}
in four variables.
A deformation that takes us to a conifold singularity is given by
\begin{equation}\label{eq:defdt}
\gamma^2 = u v w-\lambda^2(u +v +w)+2\lambda^3
\end{equation}
The conifold singularity is located at $u=v=w=\lambda$.
This 
deformation can be realized by a quantum field theory, so we can engineer
the above geometry. The field theory that realizes the above deformation is 
associated to a collection of D3-branes at
 the orbifold with discrete torsion $\BC^3/\BZ_2\times\BZ_2$.

In the case with discrete torsion, the quantum field theory for the orbifold 
$\BZ_2\times \BZ_2$ singularity consists of a quiver diagram with one gauge
 group and
three superfields $X,Y,Z$ in the adjoint representation \cite{D}.
 This is because 
the $\BZ_2\times \BZ_2$ group has only one irreducible projective 
representation.The
superpotential is given by
\begin{equation}
a\tr(XYZ+ZYX)
\end{equation}
where $a$ is a coupling constant that we set equal to one by rescaling the
fields. The field content is summarized by the quiver diagram appearing in
figure
 \ref{fig:quiver2.eps}.

\myfig{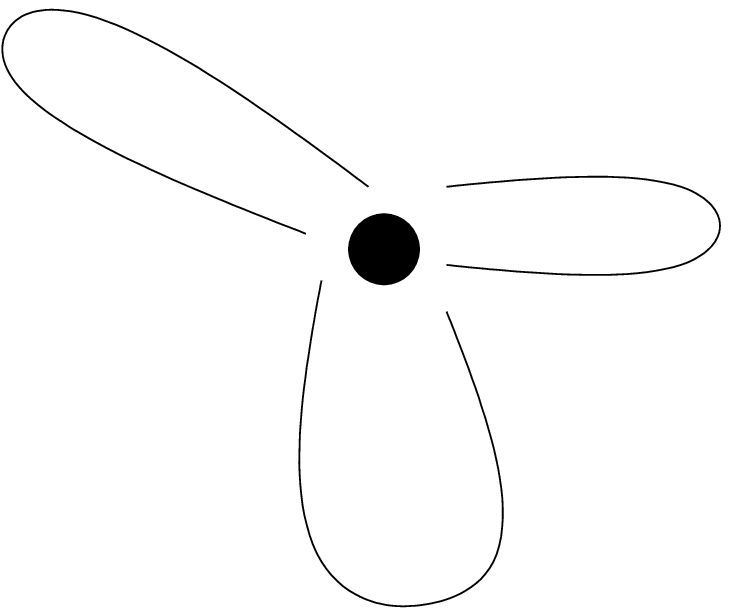}{3}{Quiver diagram for the $\BC^3/\BZ_2\times\BZ_2$
orbifold with discrete torsion}

 This field theory
was analyzed from the AdS/CFT correspondence in \cite{D, BJL} and shown
 to lead to the correct moduli
space of D-branes plus the fractional branes at the codimension two 
singularities.

The superpotential deformation that gives rise to the conifold singularity
 is obtained by
adding F-terms for the superfields \cite{D,DF}
\begin{equation}\label{eq:zeta}
-2\zeta\tr(X+Y+Z)
\end{equation}
and it is chosen so that one has a $S_3$ symmetry from permuting the three
variables $X,Y,Z$. 

We want to show now that this superpotential leads to the variety 
\ref{eq:defdt}. The modified algebra is as follows
\begin{eqnarray}
XY+YX &=& 2 \zeta \\
XZ+ZX &=& 2 \zeta \\
YZ+ZY &=& 2 \zeta
\end{eqnarray}
It is easy to verify that $u= X^2$, $v=Y^2$, $w=Z^2$ are in the center of the
algebra for any value of $\zeta$. At $\zeta=0$ also $\gamma=XYZ$ is
 in the center, but
this gets modified to
\begin{equation}\label{eq:gamma}
\gamma = XYZ - \zeta X -\zeta Z +\zeta Y = -ZYX+\zeta X +\zeta Z - \zeta Y
\end{equation}
One can now expand $\gamma^2$ by using the commutation 
relations and one finds
\begin{equation}
\gamma^2 = XYZ XYZ +\{XYZ, -\zeta X -\zeta Z +\zeta Y\}
+\zeta^2(X^2+Y^2+Z^2) -2 \zeta^3
\end{equation}
From equation \ref{eq:gamma},
 we replace $XYZ = -ZYX +2\zeta X +2\zeta Z -\zeta Y$
so the relation becomes
\begin{equation}
\gamma^2 =  - u v w +(2\zeta X+2\zeta Z - 2\zeta Y)XYZ+ 
 \{XYZ, -\zeta X -\zeta Z +\zeta Y\}
+\zeta^2(X^2+Y^2+Z^2) -2 \zeta^3
\end{equation}
which can be written in terms of a commutator
\begin{equation}
\gamma^2 =  - u v w +[(\zeta X+\zeta Z - \zeta Y),XYZ]+ 
\zeta^2(X^2+Y^2+Z^2) -2 \zeta^3
\end{equation}
but since $\gamma$ is central, we obtain from \ref{eq:gamma} that
 the commutator in the above 
expression vanishes and we are left with
\begin{equation}\label{eq:condteq}
\gamma^2 =  - u v w  
+\zeta^2(u+v+w) -2 \zeta^3
\end{equation}
which is identical to equation \ref{eq:defdt}.

For simplicity, we will  now set $\zeta = 1$.

The above algebra has one two dimensional representation for each point away 
from the singularity. Let us assume that $X^2= a^2 \neq 0$. Then,
 since $X^2$ is
in the center while $X$ is not, we find that
\begin{equation}
X = a \sigma_3
\end{equation}
From here it follows that
\begin{equation}
Y = a^{-1} \sigma_3 + b\sigma_2+c\sigma_3 
\end{equation}
With $Y^2 = a^{-2} + b^2+c^2$. We can generically 
make the choice $b\neq 0$ if $Y^2 \neq 1 /X^2$.
 Finally, we can take 
\begin{equation}
Z = a^{-1} \sigma_3 + (1-a^{-2})b^{-1}\sigma_2 
\end{equation}
We can now first 
take the limit $a\to 1$ and then take $b,c\to 0$, which shows that
all $X,Y,Z$ become identical to each other. Thus, they commute and
the representation becomes reducible into two one dimensional irreducible
 representations which are given by $X = Y= Z = \pm 1$. Notice that
the representation becomes reducible exactly at the conifold singularity
when we evaluate explicitly $\gamma, u, v, w$.

It is an easy matter to establish that for these fractional
brane configurations the D-term equations are 
satisfied and hence the D-branes fractionate at the geometric
singularity.
The representation splits into two distinct irreducibles and 
this is our first evidence that the two singularities are describing
the same universality class of four dimensional field theories, as both
have the same splitting of fractional branes at the singularity.

\subsection{Field theory for the conifold with discrete torsion}

Now we want to prove that the infrared dynamics of D-branes near the conifold
singularity (this is a point in the moduli space of vacua of the theory)
gives exactly the same low energy effective field theory as the
conifold without discrete torsion.

To do this, we need to go to the region of moduli space where the branes 
fractionate and expand the theory about that background.
Since we have a $\BZ_3$ symmetry that exchanges $X\to Y \to Z$, it is 
convenient to do a linear change variables so that the new variables have
 definite charges under this $\BZ_3$ symmetry. These will be called $A,B,C$,
 and we write
\begin{eqnarray}
X &=& A+B+C\\
Y &=& A + \eta B + \eta^2 C\\
Z &=& A + \eta^2 B + \eta C
\end{eqnarray}
Where $\eta$ is a cube root of unity.
Under the $\BZ_3$ symmetry, $A$ is neutral while $B,C$ have opposite charges.
The superpotential in these variables is given by
\begin{equation}
W = \tr[ 2(A^3 + B^3 +C^3)- 3(ABC+CBA) - 6A]
\end{equation}
And the two conifold singularity irreducible representations
corresponds to $B= C= 0$ and $A = \pm 1$.
We now want to take $N$ D-branes close to the conifold, which gives us
a 
$U(2N)$ 
theory that we  expand
in fluctuations about the conifold point in the moduli space of vacua.
Since $A=\pm1$, the gauge group is broken to $U(N)\times U(N)$ and one can 
draw a  na\"\i ve quiver from decomposing the field content under the
unbroken gauge group quantum numbers. This is depicted in figure
\ref{fig:naivquiv.eps}.

\myfig{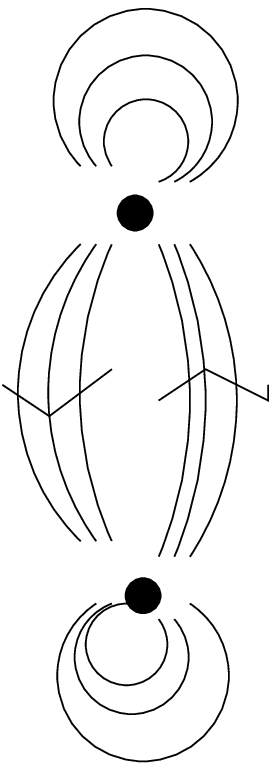}{2}{Na\"\i ve quiver diagram for the
conifold with discrete torsion}

The expansion in fluctuations is given by
\begin{eqnarray}
A &=& \begin{pmatrix}1 +\delta A_{11}& 0\\
0&-1 +\delta A_{22}
\end{pmatrix}\\
B &=&  \begin{pmatrix}\delta B_{11}& \delta B_{12}\\
\delta B_{21}&\delta B_{22}
\end{pmatrix}\\
C &=&  \begin{pmatrix}\delta C_{11}& \delta C_{12}\\
\delta C_{21}&\delta C_{22}
\end{pmatrix}
\end{eqnarray}
where we have used the allowed gauge invariance to set 
$\delta A_{12} = \delta A_{21} = 0 $. 
The equations to first order read
\begin{equation}
\{ <A>, \delta B\} = \{<A>, \delta C\} = \{ <A>, A\} =0
\end{equation}
so we find that $B_{ii}, C_{ii}, A_{ii}$ have to be zero, this is, they are 
massive fields that have to be integrated out. The massless fields are
given by the elements of the $B_{ij}$ and $C_{ij}$ matrices with $i\neq j$.
As such, we remove them from the na\"\i ve quiver diagram, as well as 
the arrows corresponding to $\delta A_{12} , \delta A_{21}$ which are the 
goldstone modes that are eaten by the Higgs mechanism.

The associated quiver diagram of the singularity is then given by two
$U(N)$ gauge groups plus two fields in the $(\bar N, N)$ and two fields in 
the $(N, \bar N)$, and this begins to look like the standard conifold 
\cite{KW}, as described by figure \ref{fig:quiver.eps}.
To find the effective superpotential, we have to integrate out the massive 
fields $A_{ii}, B_{ii}, C_{ii}$. To do this, it is necessary to expand the 
equations to second order in fluctuations. 
From here, we find that
\begin{eqnarray}
\begin{pmatrix}
\delta B_{11}^{(2)}&0\\
0& -\delta B_{22}^{(2)}
\end{pmatrix}
&=&  \begin{pmatrix}
\delta C_{12} \delta C_{21} &0\\
0& \delta C_{21}\delta C_{12}
\end{pmatrix}\\
\begin{pmatrix}\delta C_{11}^{(2)}&0\\
0& -\delta C_{22}^{(2)}
\end{pmatrix}&=& \begin{pmatrix}
\delta B_{12} \delta B_{21} &0\\
0& \delta B_{21}\delta B_{12}
\end{pmatrix}
\\
4 \begin{pmatrix}
\delta A_{11}^{(2)}&0\\
0& -\delta A_{22}^{(2)}
\end{pmatrix}
&=&
 \begin{pmatrix}
\delta B_{12} \delta C_{21}+\delta C_{12} \delta B_{21} &0\\
0& \delta B_{21}\delta C_{12}+\delta C_{21} \delta B_{12}
\end{pmatrix}
\end{eqnarray}

We now have to substitute these terms in the superpotential
to obtain the effective superpotential for the massless fields. 
We get a superpotential of the form
\begin{equation}
W = \tr<A> ( a \delta B_{(1)}^2 \delta C_{(1)}^2 + b \delta B_{(1)}
\delta C_{(1)}\delta B_{(1)}\delta C_{(1)})+\hbox{higher order}
\end{equation}

Although at first sight it looks like the superpotential 
is different from \ref{eq:sup1}, we have to remember that this is
only well defined up to field redefinitions. Indeed, for the number
of branes set equal to one, ($N=1$) we find no superpotential at all,
which is exactly 
the case for the conifold without discrete torsion as well.

Consider now the following commutator
\begin{eqnarray}
\zeta &=
&\left[\begin{pmatrix} 0 & \delta B_{12}\\
\delta B_{21} &0
\end{pmatrix},
\begin{pmatrix} 0 &- \lambda \delta C_{12}\\
\lambda'\delta C_{21} &0
\end{pmatrix}
 \right]= [B, C'] \\ & = &
 \begin{pmatrix} 
\lambda'\delta B_{12} \delta C_{21}+\lambda\delta C_{12} \delta B_{21}  & 0\\
0 &-(\lambda'\delta C_{21} \delta B_{12}+\lambda\delta B_{21} \delta C_{12})
\end{pmatrix}
\end{eqnarray}
where we have defined a modified  $C'$ with $\lambda, \lambda'$  
unitary. It follows that
\begin{equation}
\theta \tr <A> \zeta^2 = \theta\tr 
\left[<A> (2\lambda \lambda' (\delta B_{(1)}^2 
\delta C^2_{(1)})+(\lambda^2+{\lambda'}^2)
(\delta B_{(1)} \delta C_{(1)})^2)\right]
\end{equation}
So we can adjust the coefficients $\lambda, \lambda', \theta$ to match
$a, b$. With this slight change of variables we see that we are getting
the right form of the superpotential as in equation \ref{eq:sup1} with an 
identification of the 
variables.

The higher order terms in the superpotential are present but become 
irrelevant in the low energy field theory once we flow to the infrared fixed
 point at the conifold singularity.
The dictionary between the new variables and the previous ones\footnote{We
 denote the variables for \ref{eq:sup1} with a tilde}
is $<A> \sim \sigma$, $B\sim \tilde X$, $C'\sim \tilde Y$

Also notice that the $\BZ_3$ symmetry under which $B,C$ transformed 
becomes part of the $SU(2)\times SU(2)$ expected symmetry of the field
theory in the infrared, so there are no additional discrete symmetries in 
the theory with discrete torsion.

The higher order terms which are irrelevant in the infrared and which we 
discarded when going to the infrared fixed point
break the $SU(2)\times SU(2)$ explicitly. Thus, the symmetry of the theory in
the infrared is an enhanced global symmetry which exists only in the 
infrared fixed point field theory.

\section{Marginal deformations}\label{sec:md}

We have obtained through very different means the gauge 
theory corresponding to the conifold, and as such we should be able to 
match the marginal deformations from the two theories in the infrared 
fixed point. In particular, the gauge theory at the origin has two marginal
 deformations corresponding to the two gauge couplings in the quiver 
diagram, while the superpotential coupling remains fixed 
from the conditions for conformal invariance.

In the theory obtained from the $\BC^2/\BZ_2\times \BC$ deformation,
we clearly have two distinct gauge couplings in the UV field theory which
are also marginal. These are the two gauge couplings in the $A_1$
quiver theory. Thus, apart from some flow of couplings to the infrared 
region, we can identify the marginal deformations in the IR with marginal 
deformations in the UV finite field theory.

The operator is in this case
\begin{equation}
\int d^2 \theta \tr (\sigma W^2)
\end{equation}
with $\sigma W = W\sigma$ from the orbifold projection into the 
physical fields.

For the second case obtained from the orbifold with discrete torsion, there 
is only one gauge coupling in the UV finite field theory because
 we have only one 
gauge group.

However, the marginal operator in the infrared field theory should correspond
 to some operator in the UV. Since the operator should affect the gauge
couplings as they split at the representation, it is proportional
to $\sigma$. Remember now that we had identified
 $\sigma \sim <A>$, and that the gauge coupling
is a holomorphic function of the chiral fields,
so the smallest dimension
operator in the UV field theory which can accomplish this splitting is
\begin{equation}
{\cal O} = \int d^2 \theta \tr( A W^2)
\end{equation}
which is of dimension five. Thus, in the UV field theory, this 
would be considered an irrelevant coupling 
(in terms of the flow to the infrared) and would be discarded. However,
in the flow towards the infrared field theory, this operator acquires an 
anomalous dimension that makes it relevant at the IR fixed point, making 
it another example of a dangerously irrelevant operator \cite{LS}.

However, in the AdS/CFT correspondence one would not add such an operator 
to the theory in the UV because it would drastically change the boundary 
conditions. 

Thus, in this case, the theory flows to the conifold field theory at some 
specific value of the couplings. Classically the two coupling are
equal and since the equations of motion have a symmetry under $A\to -A$
this means that we are in the scenario where the gauge theory has an extra
$\BZ_2$ symmetry at the conifold, so this should be true quantum 
mechanically as well.

For the D-terms of the theory the same type of argument works. We
need to add terms 
to the effective action proportional to $\sigma$. But these, not being
holomorphic, require us to think of terms in the K\"ahler potential.

However there are no terms in the K\"ahler potential
 that can generate these types of $D$ terms.
From the superfield formulation, terms linear in the D-terms
of the vector field will be found in
\begin{equation}
\int d^4\theta \tr(f(\phi, e^-V \bar \phi e^V))
\end{equation}
and to first order this involves the commutators $[\bar\phi,V]$.
But these are exactly the gauge fields which are integrated out by the 
Higgs mechanism. These deformations, not being described by a superspace
integral,
necessarily
break supersymmetry.

The obstruction to being able to blow up the conifold tip has to do with
the fact that if such a blow-up were possible, then the 2-sphere representing
such a cycle would be a holomorphic submanifold which turns out to be
torsion in homology.
Such a manifold can not be K\"ahler because in a K\"aler manifold 
any complex
submanifold is minimizing in it's volume class (it is calibrated by the
K\"ahler form). 
Thus, the field theory and the topology agree. This also compares 
favourably with the results of \cite{AMG}, where a blowup of the
conifold singularity produced a global geometry which was not 
of the Calabi-Yau type.

To see that such a sphere is torsion we go back to the orbifold with discrete
 torsion. There we have three complex lines of $A_1$ 
singularities meeting at the 
origin. On each of these lines one can assume that there is a blown-down 
$\CP^1$ fibration. However, we have a monodromy of the fractional branes
\cite{BL}, 
so we would think of this fibration as one where we change the orientation
of the $\CP^1$  as we move on a circle about the origin. Such 
a $\CP^1$ can not be holomorphic because otherwise 
it would have  a  canonical orientation given by the complex 
structure.

This suggests that in the orbifolds with discrete torsion $T^6/\BZ_2\times 
\BZ_2$, which are deformed to a Calabi-Yau space with conifold singularities,
 the obstruction to the possibility of blowing up the conifold is a
global feature of the topology
and not a particular detail of the singularity itself. 
This seems to be in accordance with the results found in \cite{AMG}
and supports the idea that the local behavior near 
the singularity is universal.

To push this point of view further, let us consider the resolved conifold.
It is known that the blow-up mode of the conifold is non-normalizable
in the conifold geometry
\footnote{I thank J. Maldacena for showing me this result}. Thus, given a 
metric deformation that changes the volume of this cycle, one can show that
the metric perturbation is non-normalizable.
From this point of view,
in order to determine if we are allowed to blow-up the conifold singularity
 we need to 
specify boundary conditions at infinity for this mode. As such, this
 mode is sensitive
to the details of how the conifold is embedded in the Calabi-Yau space and
 can be
 expected to be  dependent on the topological features at infinity.

\section{Duality cascades, multi-trace operators and 
deformed moduli spaces.}\label{sec:cascade}

In the conifold geometry we expect to be
able to describe the duality cascades \cite{KS} in any UV description of the 
theory ,  and to generate as well a non-perturbative effective 
superpotential for brane probes.

The dualities are triggered from the effective gauge coupling becoming infinite
for  one of the fractional
branes at a given scale $\mu$. One would like a geometric interpretation of 
this same singularity in an associated system from where we can draw 
conclusion as 
to the nature of this duality. 

The D-brane system under T-dualities is dual to a matrix model of $D0$
 branes,
 where the gauge
groups $U(N)\times U(M)$ count mutually BPS fractional branes. Since the model
is associated with a deformation of an $A_1$ singularity, we will do
 the 
duality
analysis in the matrix model for the ALE space instead, which is a considerable
simplification.

Our point of view is to concentrate only on the role that the gauge coupling 
becoming infinite plays. Indeed, in the matrix model the two gauge couplings
$1/g_1^2$ and $1/g_2^2$ are real parameters that are related to Wilson lines
in M-theory, see for example \cite{BC}. The mass of a full D0-brane
is 
proportional to 
$1/g_1^2 +1/g_2^2$, which is a bound state of two fractional branes that
are mutually BPS to the D0 brane if there are no F-terms or D-terms in the 
Lagrangian, 
while the masses of the fractional branes are proportional to 
$1/g_i^2$.

The fractional branes have to be chosen so that each of them has
 positive D0
 brane charge, 
however, when we move in the gauge coupling moduli space we pass
 through a point in M-theory
where the Wilson line around the eleventh direction vanishes and one
 of the
 fractional D-branes 
becomes massless. This is 
exactly the point at infinite gauge coupling, but we are allowed to continue
past this point as the moduli space of Wilson lines does not stop here.

The D-brane charge lattice has two charges, the D0-brane charge and
 the
 fractional
D-brane charge. This is represented in the  figure \ref{fig:lattice.eps}.

\myfig{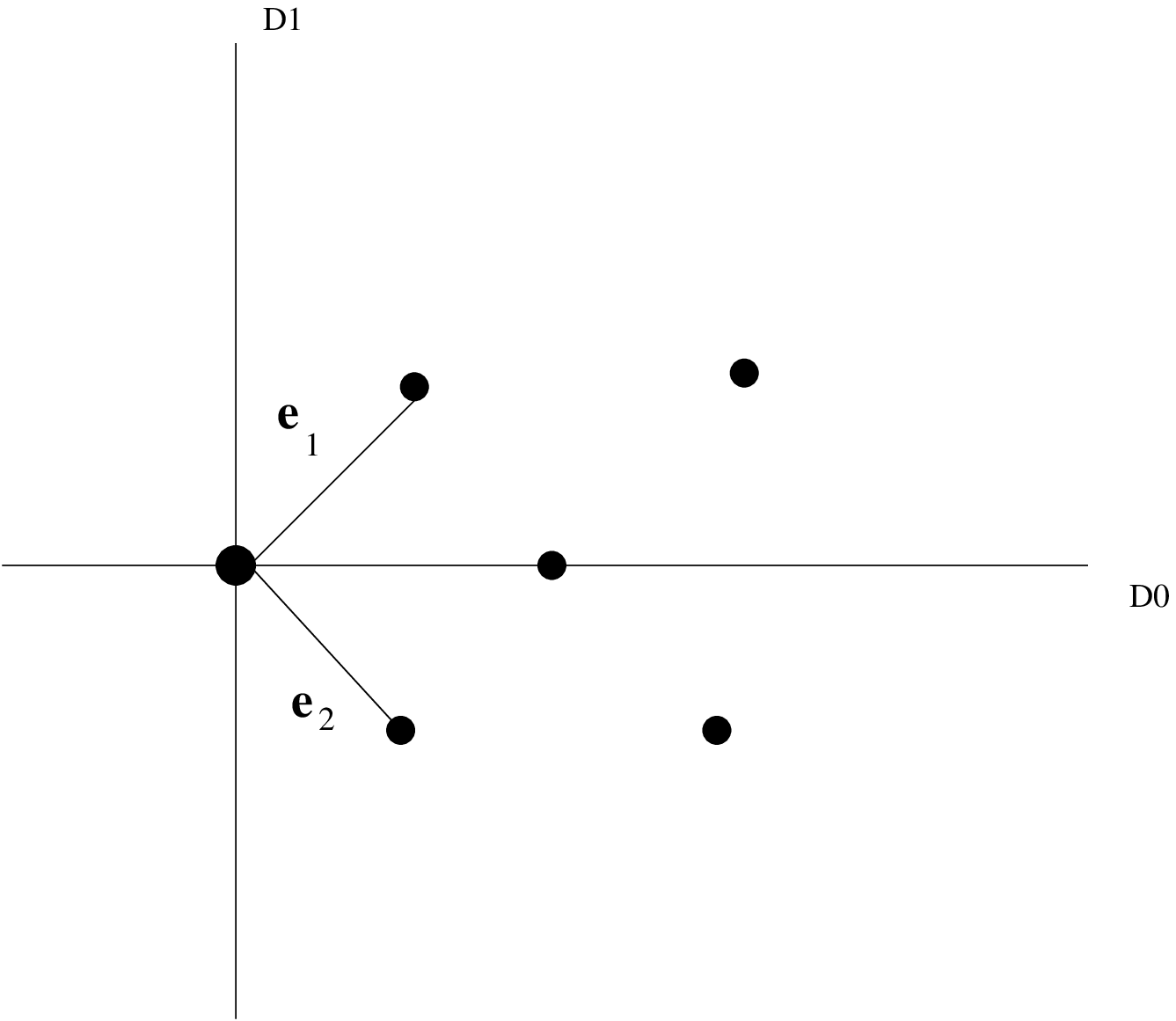}{5}{D-brane charge lattice for the $A_1$ singularity}

The way to choose the constituent fractional D-brane charges for the matrix
 model
 is to take
those  
D-branes with the smallest possible {\em positive} D0 brane charge.
When the Wilson line goes through zero and becomes negative
we need to change variables to keep our 
prescription intact, and this corresponds to a change of basis in the D-brane 
lattice.

The appropriate new lattice genertors are given by
\begin{equation}
e'_2 =  2e_1 + e_2, e'_1=-e_1 
\end{equation}
The lattice point $Ne_1+Me_2$ in terms of the new variables is given 
by $(2M-N)e_1'+ M e_2'$, and thus the new gauge group for the matrix
model
 should be given 
by $U(2M-N) \times U(M)$, with the same quiver diagram.

We recognize this immediately as the Seiberg duality \cite{Sei}
transformation
 that
 appears in the duality cascade.
 As far as the enhanced $SU(2)$ gauge model of
 M-theory, the change
of basis $e_1\to e_1'=-e_1$ on the root lattice of $SU(2)$ is a Weyl
 reflection,
a fact  noticed in \cite{CKV}. 
These changes by Weyl reflections are our choice as to which roots of the 
charge 
lattice are
declared to be simple.
In the matrix prescription mentioned above,
this choice is given by which roots have the minimal positive $D0$ brane 
charge.

This toy model helps us understand why we have Seiberg Dualities in the
 field theory.
However, notice that as far as M-theory is concerned we have to cross a 
singular point where
we have extra massless fields, whereas in the conifold flow there is no
 such phenomena
happening and the supergravity solution is nonsingular.

The difference between the conifold and the $A_1$ singularity can be understood
 because
the masses of these fractional branes also receive a contribution from the
deformation of the $A_1$ singularity. We are now in four dimensions and not
 seven 
dimensions, so the $A_1$ enhanced symmetry point is broken by
values of fields that depend on positions that are transverse to the 
four dimensions. 
However, the treatment of these moduli is not democratic.
 Some of them are 
determined
by the complex structure of the Calabi-Yau manifold, while the gauge coupling
 comes
from K\"ahler moduli. As far as the D-branes are concerned, the contributions
 to the
central charge of a D-brane are determined by the Kh\"aler moduli. Because 
of the extra
deformation the would-be fractional D-branes are not BPS at this 
position; they are massive, but their local contribution to the central charge 
vanishes.
This is depicted in figure \ref{fig:contour.eps}, 
where we miss
the singular point by taking a contour that takes into account that
 other fields than the gauge coupling  are present.

\myfig{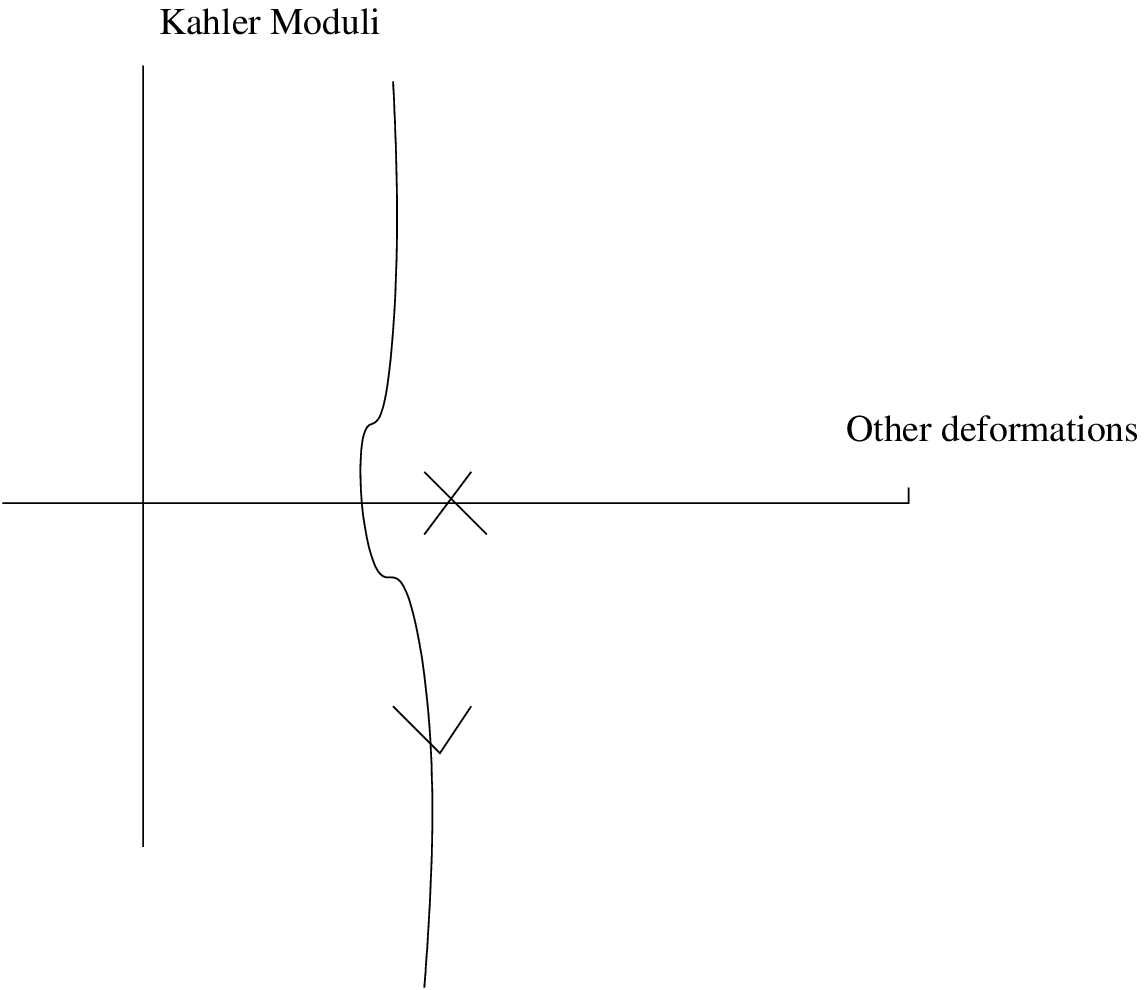}{5}{Deformed contour in closed string moduli space:
the horizontal line represents the singularity in K\"ahler moduli space. The only real
singularity is marked with the $X$.}

We still want to keep a description where the local central
 charge of a fractional object contributes in the same direction as the 
global D3 brane
 charge. To ensure this, we have to change our notion of constituent 
fractional D-branes 
as we flow towards the infrared. As the contour in figure 
\ref{fig:contour.eps} shows,
we have missed the singularity, but we still crossed the 
wall in K\"ahler moduli space, so we have an analytic 
continuation beyond the would-be singularity. This is also reminiscent 
of how 
brane crossing in D-brane setups  give rise to Seiberg dualities
\cite{EGK}.
 The change in description in crossing the 
wall is the field theory duality. This cascade
 has been also generalized to 
$SO/SP$ dualities by including orientifolds \cite{EOT}.

\subsection{Additional evidence for the deformed conifold}

In the paper of Klebanov and Strassler \cite{KS} it was proposed that
 at the
 end point 
of the duality cascades the singularity of the supergravity solution
 is
 resolved
 by a  deformed conifold.

For this interpretation to make sense, one expects that this is the
 effective
 geometry that any
number of D3-branes see. They proved that one probe D3-brane leads to
 a deformed conifold, and we want
 to move this result forward for a collection of $n$ D3 brane probes.

One expects that this moduli space is essentially the symmetric
 product of $n$ copies of the
conifold geometry $\cal M$, namely one would want to find that the
 moduli
 space of vacua gives rise
to $Sym^n({\cal M})$. We will prove this result in what follows.

Consider the theory for $n$ D3 brane probes in the background of $M$
 fractional branes. 
Then we have a theory with gauge group $U(n)\times U(M+n)$, and the
 $U(n+M)$ becomes
strongly coupled with a dynamical scale $\Lambda$.

Remember the definition of the variables $N_{ij}$ in equation \ref{eq:Nij}. 
The superpotential plus the nonperturbative correction from the
Affleck-Dine-Seiberg superpotential \cite{ADS} in these variables is given by
\begin{equation}\label{eq:quantsup}
W = \tr N_{ij} N_{kl} \epsilon^{ik}\epsilon^{jl}+
c\left(\Lambda^a/Det[N_{ij}]\right)^b
\end{equation}
with $a,b,c$ constants depending on $n,M$ and with 
\begin{equation}
\det[N_{ij}] = \det
\begin{pmatrix}
N_{11} & N_{12}\\
N_{12} & N_{22}
\end{pmatrix}
\end{equation} 
a $2n\times 2n$ matrix transforming as $(2\times2)$ matrices 
of $SU(n)$ adjoints.
The variation of the above equation leads to 
\begin{equation}
{\delta W}
= \tr[2 N_{kl} \epsilon^{ik}\epsilon^{jl} \delta N_{ij}]+
 bc\left(\Lambda^a/Det[N_{ij}]\right)^b\tr[ (N ^{-1})^{ij}\delta N_{ij}] 
\end{equation}
This is
\begin{equation}
{\delta W}
= 2 N_{kl} \epsilon^{ik}\epsilon^{jl} -
 bc\left(\Lambda^a/Det[N_{ij}]\right)^b (N ^{-1})^{ij}
\end{equation}
with $N^{-1}$ the inverse of the $2n\times 2n$  matrix $N_{ij}$.
 Now notice that the term
with the determinant has no $SU(n)$ or flavor indices, so it can be
 treated 
as a constant and we can 
rewrite the equation we need to solve as 
\begin{equation}
N_{kl} \epsilon^{ik}\epsilon^{jl} - \Omega (N^{-1})^{ij}
\end{equation}
for some constant $\Omega$.
Multiplying by $N$ on the right, we obtain
\begin{equation}
2 N_{kl} \epsilon^{ik}\epsilon^{jl} N_{jl} = \Omega \delta^i_l
\end{equation}
and a similar expression for matrix multiplication by 
$N$ on the left. The content of the above equation is
\begin{equation}
\begin{pmatrix}
N_{22}  &- N_{21}\\
-N_{12} & N_{11}
\end{pmatrix} 
\begin{pmatrix}
N_{11} & N_{12}\\
N_{21} & N_{22}
\end{pmatrix}
= 
\begin{pmatrix}
\Omega & 0\\
0 & \Omega
\end{pmatrix} = \begin{pmatrix}
N_{11} & N_{12}\\
N_{21} & N_{22}
\end{pmatrix}\begin{pmatrix}
N_{22}  &- N_{21}\\
-N_{12} & N_{11}
\end{pmatrix} 
\end{equation}
The above equations imply that $N_{12}$ and $N_{21}$
commute with $N_{11}$ and $N_{22}$. 
If $N_{11}$ is generic, we expect that it can be diagonalized with
 distinct eigenvalues
by a complexified gauge transformation. Since $N_{12}$ and $N_{21}$ 
commute with $N_{11}$, 
it follows that they are diagonalized by the same basis. Then 
$[N_{12},N_{21}] = 
[N_{11}, N_{22}] = 0$ so that all of the $N_{ij}$ matrices commute 
with one another, and the
collection of matrices $N_{ij}$ in the adjoint of $SU(n)$ are  simultaneously 
diagonalizable. 

The constraint satisfied by the eigenvalues is exactly the equation
corresponding to the deformed 
conifold, so the moduli space of vacua 
is given by $N$ copies of the deformed conifold geometry up to $SU(N)$
 gauge transformations
which permute the eigenvalues. This is, the moduli space of vacua
 is the symmetric product of the conifold
space geometry.

All we need to do now is match the scale $\Lambda$, or equivalently
$\Omega$, to the deformation parameter and we find exact agreement with
the proposal of \cite{KS} to add an arbitrary number of probes to the geometry.

This is a strong test of the deformed conifold proposal, as the above
 result could have been
expected to hold only in the semiclassical regime where the vevs of
 all
 of the $N$ are very large and 
to receive large  quantum deformations when the scale associated to
 the vevs $<N>$ are small.

There are still singularities in the moduli space where two of these probe 
D3-branes meet. The low energy effective
field theory degrees of freedom are such that one expects and enhanced $SU(2)$
gauge theory. The field counting 
suggests that these have an effective $N=4$ supersymmetry so that these
singularities are not resolved.

\subsection{More on discrete torsion}

The statements made in the previous subsection depend only on
effective field theory. However, one would want an ultraviolet
description of the same statements in the theory with discrete 
torsion. Now we will try to do some consistency checks in this case 
to make sure that the geometrical description is accurate.

First, it is expected that we will deform the geometry without
affecting the structure of the manifold at infinity. The constraints 
in equation \ref{eq:defdt}
are modified to quantum constraints
\begin{equation}
\gamma^2 = u v w-\lambda^2(u +v +w)+2\lambda^3+\epsilon
\end{equation}
with $\epsilon$ being determined by the gaugino condensate.
This becomes regular everywhere and it is the most
economical possibility for the deformed quantum potential.

What is interesting in this case is that the constants 
$a,b,c$ appearing in equation \ref{eq:quantsup} depend on 
$n,M$, the number
of total branes and the number of fractional branes.
In our case the number of fractional branes is counted
by the superfield vev $\tr(A)$, and this is determined dynamically by
the vacuum choice, as opposed to being imposed by hand as a superselection 
sector.

The superpotential might look like
\begin{equation}\label{eq:weff2}
W_{tree} + a(\tr(A)) \left(\Lambda^b(\tr(A))/\det(N)\right)^c(\tr (A))
\end{equation}
which looks like a non-analytic function of the chiral fields and this
might be considered pathological. However, the classical vev of
$\tr(A)$ is quantized so these worries might be unfounded. The point
is that, if we are trying to do a variation of the above equation, we
need to worry about the vacuum structure even if there are no
fractional branes because all the vacua of the theory 
with and without fractional branes are connected to one another.
The only part which is difficult to trace is the variation with
respect to $A$, but notice that in equation \ref{eq:weff2} we can vary $A$
assuming $\det(N)$ is constant. After all this is just one function on the
moduli space of $n$ branes, and not a whole matrix variable, so we can 
treat is a constant matrix proportional to the identity. The same is
true
for values of $\tr(A)$ obtained after the variation of the superpotential.
In the end, this variation might at worse renormalize the value of $\zeta$
\ref{eq:zeta}, but one does not even expect this to happen unless
$\tr(A)\neq 0$

The non-perturbative superpontential should vanish for 
$\tr(A)=0$ which is the condition for no fractional branes. This can
be accomplished if there is a $\tr(A)^2$ prefactor in the
nonperturbative
terms of the superpotential. Since the theory is well defined in the
UV with a given gauge coupling and value of $\zeta$, the
nonperturbative scale
$\Lambda$ can be written in terms of these UV quantities, and notice
that $\zeta^{1/2}  = \tr(A)$ for any  of the two
fractional brane solutions as well.

As such, these corrections will all be proportional to $\tr(A)$
which vanishes for any configuration  without 
fractional branes, and we recover our undeformed conifold
singularity when we expect it to be there. 

Also, if $\tr(A)\neq 0$, then we get nonperturbative corrections to the 
geometry, and the low energy effective field theory near the conifold
predicts that we obtain a deformed conifold.

Notice that the nonperturbative form of the 
 superpotential is of multi-trace type 
generically (the determinant can also be written in terms of multiple
traces). Na\"\i vely this seems to produce exotic non-localities
between D-branes \cite{ABS}, but in this case it seems that rather mysteriously
 we have obtained a very well defined 
local interpretation of the resulting geometry in terms of a 
deformed conifold.

\section{Holography and universality}\label{sec:hol}

We want to argue that the phenomenon encountered previously for the conifold 
is generic:
given a singularity in a Calabi-Yau manifold that is associated to a 
nonsingular worldsheet 
${\cal N}=(2,2)$ sigma model in type II theories, without any $RR$ gauge fields
and without $NS$ five-branes at the singularity,
the classes of 
field theories associated to D-brane probes on the cone geometry are
universal (up to Seiberg dualities). This is clear for singularities on a 
Calabi-Yau two-fold. They are all given by quotient singularities, and the
resolutions of singularities by blow-ups are unique giving rise to an 
ADE Dynkin diagram for the exceptional divisors.
This universal behavior has been found 
in toric singularities \cite{FHH,FHH2,BPR,FHHU} and in the Weyl
reflections for $ADE$ fibrations in \cite{CKV, CIKV}.
The argument we present is speculative and should be considered more 
as a hint of universality rather than a proof.
 As such one should approach this 
section as establishing some evidence for a conjecture.

In order to describe this conjecture in more detail, we first 
need to give the data for a singularity. If a singularity $\CM$ is isolated
(there are no codimension two singularities that pass through the codimension
 three singularity), then the base of the cone over the singularity, which we 
will label as $X$ is a smooth Sasakian 
 manifold, see \cite{MP}. Let us also assume that $X$ has no torsion
homology.

If we put a set of D3-branes near the singularity then we will get a near 
horizon geometry of the form $AdS_5\times X$. The moduli space of vacua of 
D3-branes should be given by the symmetric product $Sym^n(\CM)$, and from 
the ideas of \cite{KW2} one can actually measure this moduli space in
 terms of boundary conditions for the supergravity in $AdS_5\times X$.
From the geometry of $X$ one can also determine the chiral ring 
of the associated large $N$ field theory completely, as well as the 
global symmetries of the field theory. 

This data is universal regardless of what field theory realizes the 
singularity. Usually, given two field theories which have the same moduli 
space of vacua, the same global symmetries and the same chiral ring they
can be dual to one another in the infrared \cite{Sei}.

If the singularity is non-isolated we need to be more careful. 
This is because the supergravity approximation breaks down, as the
singularities meet $X$ on a circle. These codimension two singularities 
are going to be of ADE type on $X$. Thus, it is clear locally how the 
resolutions look like from analyzing the twisted sector strings. 
However, as these are on a circle, there is a possibility of having
monodromies for the blow-up modes. These monodromies are 
given by the choices of  discrete torsion on orbifolds \cite{BL}, and there
are consistency conditions that correlate the different monodromies on 
distinct circles. Presumably these consistency conditions are the ones
 that guarantee that the closed string field theory without the D-branes is
 non-singular. 

The monodromies are part of the data that can be measured away from the 
singularity,  the moduli space of D-branes and the chiral ring
are sensitive to these 
monodromies, so these are holomorphic observables as well. If we give the 
additional data
of these monodromies, then we have specified the topological type of the 
singularity. Again, any two field theories with the same moduli space of 
vacua, chiral ring and global symmetries associated to the same singularity 
should be dual to one another.

The idea for our universality argument is that  in the AdS/CFT correspondence
the codimension three singularity disappears from the near horizon geometry 
and is encoded holographically on the base of the cone $X$ at the
boundary.

There are subtleties to this idea if $X$ has torsion elements in 
$H^3(X,Z)$. In principle one can excite a topologically non-trivial
$H$ field which is flat that would correspond to a choice of 
discrete torsion. This is holographic data, but at first sight
 it is hard to find
holomorphic observables which correspond to these choices. 
Wrapping branes on these cycles
gives us the spectrum of baryon-like operators (see \cite{Wbar})
in the conformal field 
theory and due to anomalies it is 
sensitive to this data.
This additinal data is then necessary to establish the universality class of
field theories at the singularity.
In the conifold case, there is no torsion in homology, so there are no
 discrete torsion choices of this type to be made.

One would want a more constructive argument. Given a singularity, 
one would want to find a recipe for constructing a
non-commutative resolution in the sense of 
\cite{BL4}, and then one has a well defined field theory associated to 
the singularity. On this field theory one can do various changes of 
D-brane basis which should correspond to Seiberg dualities, and these can be
triggered by  motion
in the moduli space of couplings of the field theories, as in the 
duality cascades. 
One can expect that, in some sense, these dualities are triggered by crossing 
lines of marginal stability, and that they correspond to D-brane monodromies
 given by Fourier-Mukai transforms \cite{AD}.

\section{Conclusion}\label{sec:con}

A careful analysis of the field theory associated to the conifold
with discrete torsion has been presented. We saw that this field
theory in the infrared is identical to the conifold
field theory constructed by Klebanov and Witten \cite{KW}.
As such, conifold singularities seem to have no additional discrete torsion
degrees of freedom at the singularity.
Obstructions to resolving the singularity are instead given by global 
topological considerations and  depend on how the 
singularity is embedded in a particular Calabi-Yau space.

We have also presented some evidence that there are no additional 
discrete degrees of freedom at any Calabi-Yau
singularity located at a point $p$
that can not be guessed 
by data on the boundary of a finite size cone around $p$. 
This is a conjecture that states that the geometric classification of 
singularities depends on variables in the low energy 
string theory  (like monodromies of fractional branes if certain
 cycles are of zero size) and the algebraic geometric 
type of the singularity.

The following properties are conjectured
 and are applicable so long as 
the Calabi-Yau manifold is considered to be very large (non compact).
This is required because we want to guarantee that D3-branes are 
stable objects \cite{AL}, only then we can trust that our field theory on the
world volume of the D-branes is supersymmetric and that the heavy string
 modes can be integrated out.

\begin{enumerate}

\item Field theory duals of D-branes at Calabi-Yau singularities are
 uniquely determined
 by the singularity type up to Seiberg dualities \cite{Sei}.
The singularity type can include discrete torsion choices 
as either monodromies of fractional branes on 
codimension two singularities, or as torsion elements of
$H^3(X,\BZ)$ if the singularity is a cone over $X$. Both of these
data are holographic.

\item The singularities give rise to field theories 
 that can have duality cascades if one can add 
fractional branes.
The Seiberg dualities are changes of basis in the fractional
brane lattice and correspond to crossing walls in 
K\"ahler moduli space.

\item One can also cross these walls by changing the boundary conditions
at infinity (bare values of the couplings)
and induce Seiberg dualities in theories without fractional 
branes.

\item Point-like branes in the Calabi-Yau manifold
 always fractionate at singularities, and 
never fractionate away from singularities.
\end{enumerate}

One would also like to find a way of constructing field theories given 
a singularity. For toric singularities one has the proposal of 
\cite{FHH}, however, one would want a more general description 
for singularities which are non-toric.
As far as I   know, all examples of Calabi-Yau threefold
 singularities
 can be obtained from either partial resolutions
or deformations of known orbifold constructions; or by orbifolding
known singularities.
 Perhaps one can address
singularities systematically in this manner, as all of the field theories
associated to these geometric processes would be constructible.

\section*{Acknowledgements}

 I would like to thank
the ITP at Santa Barbara, the ITP workshop on M-theory,
 and the Aspen Center for Physics for their hospitality and support,
where some of the
ideas appearing in this paper were first conceived. Research supported by
DOE grant DE-FG02-90ER40542. This research was supported in part by
NSF under grant no PHY99-07949.
Ii is a pleasure to thank P. Aspinwall,
M. Douglas, 
I. Klebanov, R. G. Leigh, J. Maldacena, D. Morrison,
R. Plesser, E. Silverstein, and  E. Witten for useful conversations. 
Special thanks to J. Sonnenschein for valuable discussions and 
 for commenting on 
a preliminary version of the manuscript. I would also like to thank A. 
Berenstein
for pointing various typos in the manuscript.


\providecommand{\href}[2]{#2}\begingroup\raggedright\endgroup

\end{document}